\documentclass[aps,prl,reprint,showpacs,superscriptaddress]{revtex4-1}

\usepackage{graphicx}
\usepackage{siunitx}
\usepackage[draft]{fixme}
\usepackage{amsmath}
\usepackage{mhchem}
\usepackage{xspace}
\usepackage{hyperref}
\usepackage{microtype}

\newcommand{\BR}{\textup{BR}}
\newcommand{\Carbon}{\ce{^{12}C}\xspace}
\newcommand{\Boron}{\ce{^{12}B}\xspace}
\newcommand{\Nitro}{\ce{^{12}N}\xspace}
\newcommand{\mwidth}{\columnwidth}
\newcommand{\etal}{\textit{et al.} }
\newcommand{\unsim}{\mathord{\sim}}

\newcommand{\fex}{\SI{4439}{\keV} }
\newcommand{\hylvalue}{\SI{0.58(2)}{\%}\xspace}
\newcommand{\litvalue}{\SI{1.2(3)}{\%}\xspace}

\newcommand{\mSI}[2]{\num{#1}-\si{#2}}
\newcommand{\mfex}{\mSI{4439}{\keV} }

\begin{document}


\title{Independent measurement of the Hoyle state $\beta$ feeding from $^{12}\mathrm{B}$  using Gammasphere}


\newcommand{\Aarhus}{Department of Physics and Astronomy, Aarhus University 8000 Aarhus C,
  Denmark}
\newcommand{\ANL}{Physics Division, Argonne National Laboratory, Argonne, IL 60439, USA}
\newcommand{\York}{Department of Physics, University of York, York Y010 5DD, United Kingdom}
\newcommand{\Marshall}{Marshall Space Flight Center, Huntsville, Alabama 35811, USA}
\newcommand{\Argentina}{Laboratorio Tandar, Comisión Nacional de Energía Atómica, B1650KNA
  Buenos Aires, Argentina}
\newcommand{\Maryland}{Department of Chemistry and Biochemistry, University of Maryland,
  College Park, MD 20742, USA}
\newcommand{\Michigan}{Physics Department,Western Michigan University, Kalamazoo, MI 49008,
  USA}
\newcommand{\Chicago}{Department of Astronomy and Astrophysics, University of Chicago, Chicago,
  IL 60637, USA}
\newcommand{\Joint}{Joint Institute for Nuclear Astrophysics, Chicago, IL 60637, USA}
\newcommand{\NotreDame}{Present address: Physics Department, University of Notre Dame, Notre
  Dame, Illinois 46556, USA This}
\newcommand{\TRIUMF}{TRIUMF, 4004 Westbrook Mall, Vancouver, British Columbia V6T 2A3, Canada}
\newcommand{\LSU}{Louisiana State University, Department of Physics \& Astronomy, 224 Nicholson
  Hall, Tower Dr., Baton Rouge, LA 70803-4001, USA}
\newcommand{\FSU}{Department of Physics, The Florida State University, Tallahassee, FL 32306, USA}

\author{M. Munch} \affiliation{\Aarhus}
\author{M. Alcorta} \affiliation{\TRIUMF} \affiliation{\ANL}
\author{H. O. U. Fynbo} \email[Corresponding author: ]{fynbo@phys.au.dk} \affiliation{\Aarhus}
\author{M. Albers} \affiliation{\ANL}
\author{S. Almaraz-Calderon} \altaffiliation{Present address \FSU} \affiliation{\ANL} 
\author{M. L. Avila} \affiliation{\ANL}
\author{A. D. Ayangeakaa} \affiliation{\ANL}
\author{B. B. Back} \affiliation{\ANL}
\author{P. F. Bertone} \altaffiliation{Present address: \Marshall} \affiliation{\ANL}
 \author{P. F. F. Carnelli} \affiliation{\Argentina}
 \author{M. P. Carpenter} \affiliation{\ANL}
 \author{C. J. Chiara} \affiliation{\ANL} \affiliation{\Maryland}
 \author{J. A. Clark} \affiliation{\ANL}
 \author{B. DiGiovine} \affiliation{\ANL}
 \author{J. P. Greene} \affiliation{\ANL}
 \author{J. L. Harker} \affiliation{\ANL} \affiliation{\Maryland}
 \author{C. R. Hoffman} \affiliation{\ANL}
 \author{N. J. Hubbard} \affiliation{\York}
 \author{C. L. Jiang} \affiliation{\ANL}
 \author{O. S. Kirsebom} \affiliation{\Aarhus}
 \author{T. Lauritsen} \affiliation{\ANL}
 \author{K. L. Laursen} \affiliation{\Aarhus}
 \author{S. T. Marley} \altaffiliation{Present address: \LSU} \affiliation{\ANL} \affiliation{\Michigan}
 \author{C. Nair} \affiliation{\ANL}
 \author{O. Nusair} \affiliation{\ANL}
 \author{D. Santiago-Gonzalez} \affiliation{\ANL} \affiliation{\LSU}
 \author{J. Sethi} \affiliation{\ANL} \affiliation{\Maryland}
 \author{D. Seweryniak} \affiliation{\ANL}
 \author{R. Talwar} \affiliation{\ANL}
 \author{C. Ugalde} \affiliation{\ANL} \affiliation{\Chicago} \affiliation{\Joint}
 \author{S. Zhu} \affiliation{\ANL}


\date{\today}

\begin{abstract}
  Using an array of high-purity Compton-suppressed germanium detectors, we performed an
  independent measurement of the $\beta$-decay
  branching ratio from \Boron to the second-excited state, also known as the Hoyle state, in \Carbon. Our result is
  \SI{0.64(11)}{\%}, which is a factor $\unsim 2$
  smaller than the previously established literature value, but is in agreement with another
  recent measurement. This could indicate that the Hoyle state is more clustered than
  previously believed.
  The angular correlation of the Hoyle state $\gamma$ cascade
  has also been measured for the first time. It is consistent with theoretical predictions.
\end{abstract}


\maketitle

\section{Introduction}

Carbon is the fourth most abundant element in the Universe and it plays a key role in stellar
nucleosynthesis. It is mainly formed in stars at a temperature of
$10^{8}$-$10^{9}$\si{\K}
in the triple-$\alpha$
fusion reaction, which proceeds via the second-excited state, also known as the Hoyle state, at
\SI{7.65}{\MeV} in \Carbon, famously proposed by Hoyle in 1953 \cite{Hoyle1954}.

The first attempt to theoretically explain the structure of the state was the linear alpha
chain model by Morinaga in 1956 \cite{Morinaga1956}, where he, furthermore, conjectured a $2^{+}$
state in the \SIrange{9}{10}{\MeV} region. Several more sophisticated models have been developed
since, as summarized in Ref. \cite{Freer2014}. Most of these models predict a collective $2^{+}$
excitation of the Hoyle state in the region of \SIrange{0.8}{2.3}{\MeV} above it.
Interestingly, the collective state increases the triple-$\alpha$ reaction rate at $T > 10^{9}\,\si{\K}$
by a factor of 5-10 compared to the results of Caughlan \etal \cite{Caughlan1988,Freer2009}. This makes
it highly relevant for core collapse supernovae \cite{Tur2007,Tur2010,Magkotsios2011,The1998}.

Experimentally, it is challenging to investigate this energy region, since there are
contributions from several broad states and from the so-called Hoyle state ``ghost anomaly''
\cite{Barker1962,Wilkinson1963,Selove1990}. Using inelastic proton scattering, Freer \etal provided the
first evidence for a broad $2^{+}$
contribution at \SI{9.6(1)}{\MeV} with a width of \SI{600(100)}{\keV} \cite{Freer2009}. Itoh
\etal corroborated these results using inelastic $\alpha$-scattering
\cite{Itoh2011} and a simultaneous analysis was published as well \cite{Freer2012}.  Results from an
experiment using the alternative $\Carbon(\gamma,\alpha)\ce{^{8}Be}$
reaction also identified a $2^{+}$
state in this region, but at $10.13_{-0.05}^{+0.06}$\si{\MeV}
and with a much larger width of $2080^{+330}_{-260}$\si{\keV}
\cite{Zimmerman2013,Zimmerman2013phd}. The reason for this discrepancy is presently not
understood. A natural explanation would be that several $2^{+}$ resonances are present in the
region, and that the different reaction mechanisms populate these with  different strength. 

An alternative experimental probe is the $\beta$
decay of \Boron and \Nitro. Due to the selection rules, decay of these $1^{+}$
systems will predominantly populate states with spin and parity $0^{+}$,
$1^{+}$
or $2^{+}$
and not the $3^{-}$
state at \SI{9.64}{\MeV}, which is the dominant channel in inelastic scattering
experiments. This technique has been used in several studies of \Carbon
\cite{Cook1957,Alburger1977,Fynbo2005,Diget2005,Hyldegaard2009,Hyldegaard2010}, but none of these has
identified a $2^{+}$ state at \SI{10}{\MeV}. The $\beta$ decay populates a somewhat featureless
excitation spectrum in \ce{^{12}C} which is analyzed with the R-matrix formalism in Ref. \cite{Hyldegaard2010}.     
This analysis identified both $0^{+}$ and $2^{+}$ resonances in the \SI{10.5}{\MeV} to \SI{12}{\MeV} region
with recommended energies for both resonances at \SI{11}{\MeV}. The R-matrix analysis includes a
large contribution from the high energy tail of the Hoyle state, which is sometimes referred to as
the ``ghost anomaly'' \cite{Barker1962,Wilkinson1963}. This contribution is strongly dependent
on the branching ratio with which the Hoyle state is populated in the $\beta$ decay. 

In the most recent experimental study of the $\beta$ decay, the beam was implanted in a silicon detector, which provided accurate
normalization of the branching ratios, resulting in a revision of several of these
\cite{Hyldegaard2009}. More specifically, the branching ratio to the Hoyle state from the decay
of \Boron was determined to be \hylvalue, which is inconsistent with the previously established
value of \litvalue \cite{Selove1990,Hyldegaard2009b} (\SI{1.5(3)}{\%} is listed in Ref. 
\cite{Selove1990}, but this should be revised \cite{Hyldegaard2009b}). The reduced branching
ratio for the population of the Hoyle state was used in the R-matrix analysis
\cite{Hyldegaard2010}. Furthermore, as the $\beta$ decay to a pure $3\alpha$-cluster system is forbidden, a precise measurement of the branching ratio will provide insight into the
strength of the cluster-breaking component of the Hoyle state \cite{Kanada-Enyo2007}. It is
therefore important to provide experimental confirmation of the reduced branching ratio
measured in Ref. \cite{Hyldegaard2009}.

Here, we report on an independent measurement of this branching ratio through a measurement of the
$\gamma$ decay of the Hoyle state with an array of high-purity germanium detectors. The results of a
preliminary analysis have been reported in \cite{Alcorta2014}.

\section{Method}

\begin{figure}[tb]
  \centering
  \includegraphics[width=0.935\columnwidth]{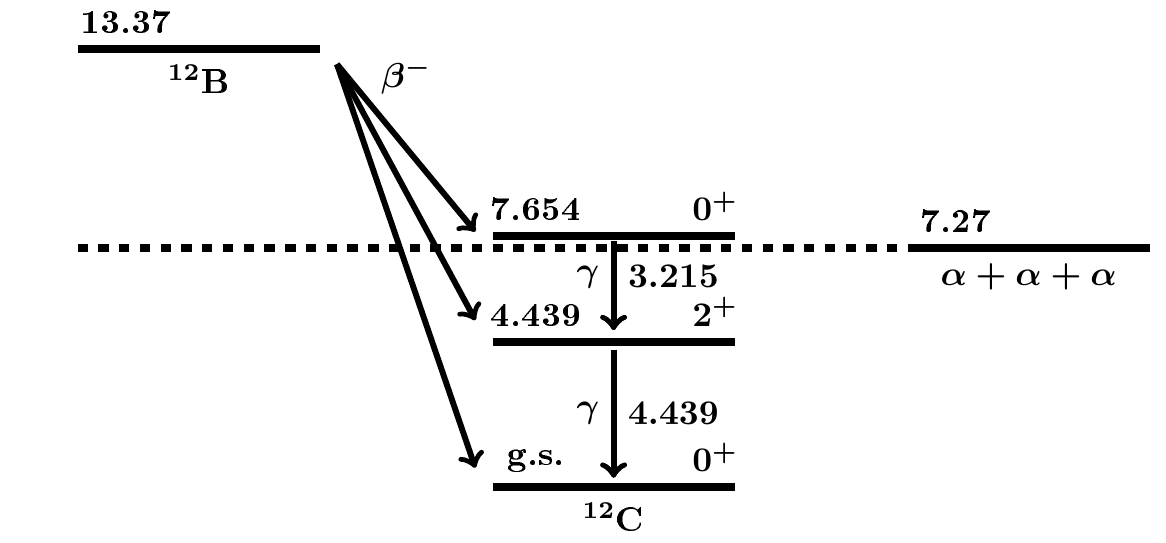}
  \caption{Level scheme of \Carbon also showing the $\alpha$-threshold and the \Boron
    ground state. Energies are given in \si{\MeV} relative to the ground state of \Carbon \cite{Selove1990}.}
  \label{fig:level_scheme}
\end{figure}

Figure~\ref{fig:level_scheme} shows the lowest states in \Carbon, the triple-$\alpha$
threshold and the ground state of \Boron. The first-excited state is below the
$\alpha$ threshold
and will only $\gamma$ decay,
whereas the Hoyle state cannot $\gamma$ decay
directly to the ground state as it is a $0^{+}$
level. It can, however, decay via the first-excited state by emission of a \mSI{3215}{\keV}
photon.

The number of $\gamma$ decays from the Hoyle state can be determined by counting the number of
\mSI{3215}{\keV} photons, and by furthermore requiring a simultaneous detection of a
\mfex $\gamma$ ray, the
background is vastly reduced. The product of the branching ratio to
the Hoyle state and its relative $\gamma$ width can then be determined by normalizing to the decay of
the first-excited state
\begin{equation}
  \label{eq:br}
  \BR(7.65) \frac{\Gamma_{\gamma}}{\Gamma} = \BR(4.44) \frac{ N_{\gamma \gamma}}{N_{4.44} \epsilon_{3.21} C_{\theta}}, 
\end{equation}
where $N_{\gamma \gamma}$ is the number of coincidence events, $\epsilon_{3.21}$ the efficiency for detecting a
\mSI{3215}{\keV} photon and $C_{\theta}$ corrects for the angular correlation between the two photons. 

The relative $\gamma$
width can be determined from all available data for the relative radiative width
\cite{Obst1976}, excluding Seeger \etal \cite{Seeger1963}, by subtracting the recommended
relative pair width from \cite{Freer2014}; yielding
$\frac{\Gamma_{\gamma}}{\Gamma} = \num{4.07(11)E-4}$.
A conservative estimate of the branching ratio to the first excited state,
$\BR(4.44) = \SI{1.23(5)}{\%}$ has been published in \cite{Selove1990}.

Using this method, the branching ratio can be determined solely with $\gamma$-ray
detectors, providing an experiment with entirely different systematic uncertainties than
previous measurements based on detection of $\alpha$ or $\beta$ particles.

\section{Experiment}

\begin{figure}[t]
  \centering
  \includegraphics[width=0.466\columnwidth]{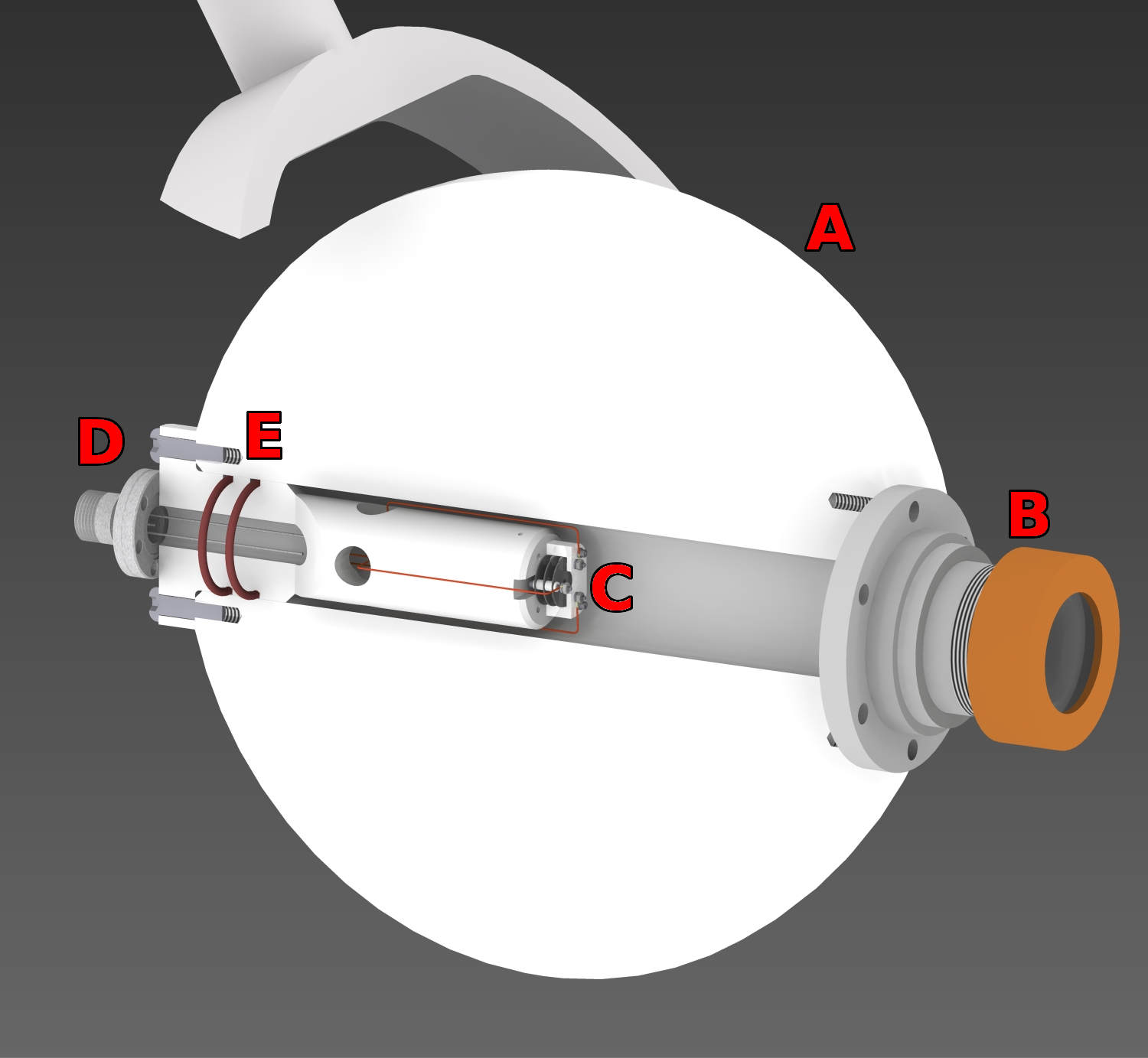}
  \caption{(Color online) CAD drawing of the chamber manufactured from a Bonner
    sphere. (A) Bonner sphere, (B) vacuum flange, (C) target holder / Faraday cup, (D)
    electrical feed through, (E) O-rings}
  \label{fig:bonner}
\end{figure}

\Boron was produced via the $\ce{^{11}B}(d, p)\Boron$
reaction in inverse kinematics, using a pulsed (\SI{40}{\ms} on, \SI{40}{\ms} off)
\mSI{40}{\MeV} \ce{^{11}B} beam delivered by the Argonne Tandem-Linac Accelerator System (ATLAS)
located at Argonne National Laboratory. A deuterated titanium foil (\ce{TiD_2}), sufficiently
thick to stop the beam, was used as target. The target was manufactured according to
the method discussed in Ref.  \cite{Greene2010} and it contained approximately
\SI{1.5}{\mg\per\cm^2} deuterium (estimated by weight).

Photons were detected using Gammasphere \cite{Lee1990}, which is an array of 110 high-purity
Compton-suppressed germanium detectors of which 98 were operational during the experiment. The
array was operated in singles mode, where any of the detectors could trigger the data
acquisition (DAQ). Data were only acquired during the beam-off period. Therefore, only delayed
activity was measured (the half life of \Boron is \SI{20.20(2)}{\ms} \cite{Selove1990}). For
each event, the time relative to beam-off as well as the energy and time for each $\gamma$
ray in the detectors were recorded.

In order to minimize bremsstrahlung caused by high-energy $\beta$ particles,
a low-Z chamber was designed - see Figure~\ref{fig:bonner}. The chamber was manufactured from a
Bonner sphere and was designed to minimize contribution from bremsstrahlung while
maintaining high gamma-ray efficiency.

\section{Analysis}

\subsection{Yield}

\begin{figure}[tb]
  \centering
  \includegraphics[width=\mwidth, clip=true, trim=0 0 0 5mm]{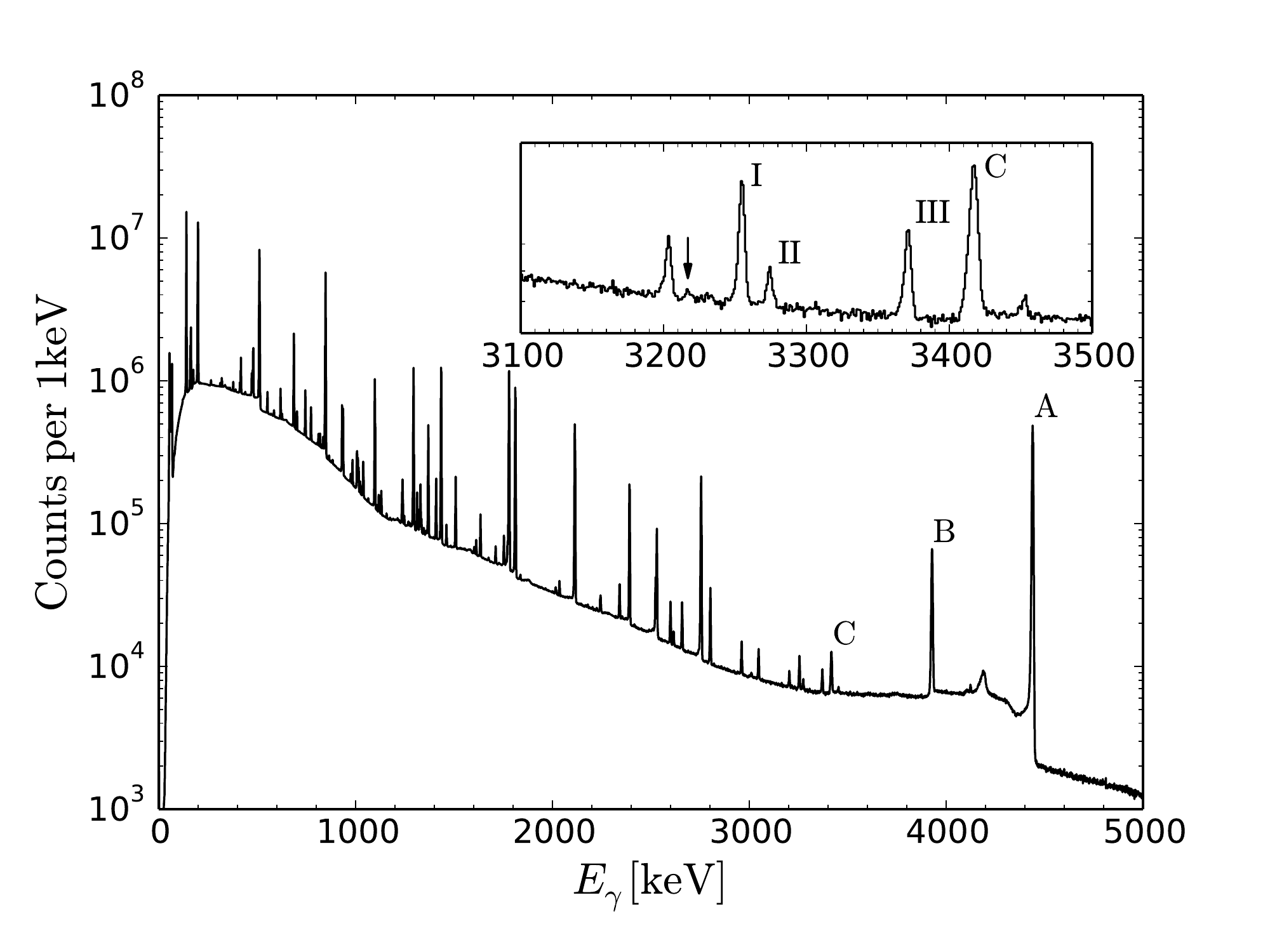}
  \caption{The entire singles spectrum acquired during the beam-off period. The \mfex peak (A) and its escape
    peaks (B,C) are clearly visible. The insert shows the \SIrange{3.1}{3.5}{\MeV} region. A small
    structure is visible around \SI{3215}{\keV}, indicated by an arrow.}
  \label{fig:full_spec}
\end{figure}

During the experiment $\unsim 10^9$
$\gamma$
rays were collected in 67 hours. The singles spectrum is displayed in Fig.~\ref{fig:full_spec},
where the transition from the first-excited state in \Carbon at
\SI[parse-numbers=false]{4439.5 \pm 0.7 (sys)}{\keV} (A) together with the first (B) and
second escape (C) peaks at lower energy are clearly seen. The insert shows the region from
\SIrange{3.1}{3.5}{\MeV} in which a structure around \SI{3215}{\keV} is visible, as indicated
by an arrow. However, the region is dominated by a peak at \SI{3200}{\keV}.

The \mfex peak was fitted with a sum of a Gaussian distribution, a skewed Gaussian
distribution, a linear background and a smoothed step function \cite{Debertin1988}.  In order
to minimize systematic effects, the fit was performed with the Poisson log likelihood ratio
\cite{Bergmann2002} using the \textsc{minuit} minimizer~\cite{James1975}. From this procedure, the area
of the peak was determined to be $N_{4.44} = \num{9.20(2)E6}$,
where the error was dominated by uncertainties in the functional form of the peak.

\subsection{Coincidence spectrum}
\label{sec:coin_spec}

\begin{figure}[t]
  \centering
  \includegraphics[width=\mwidth, clip=true, trim=0 0 0 5mm]{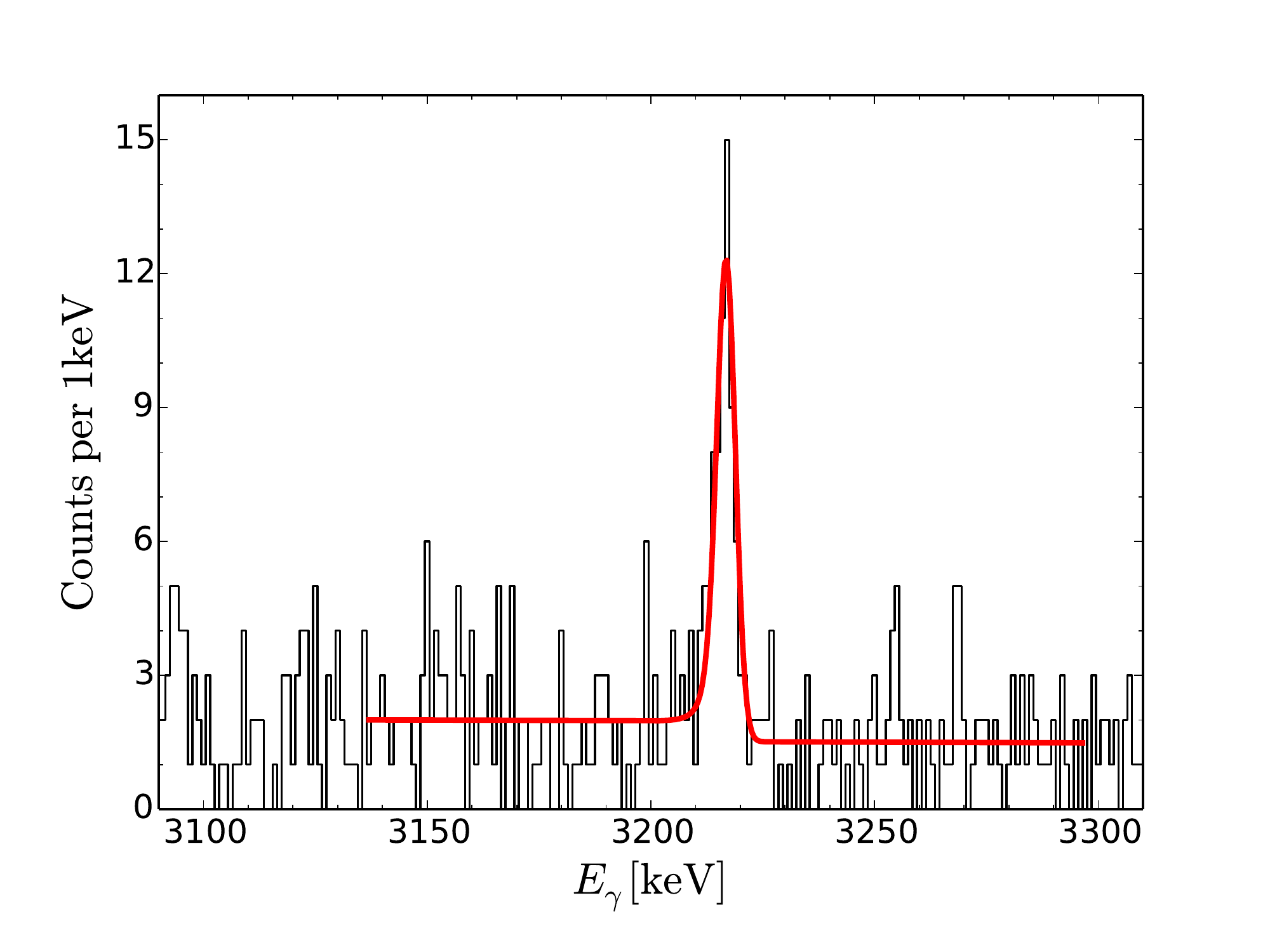}
  \caption{(Color online) Coincidence spectrum, acquired by gating on the \fex peak and the
    time difference. A clear peak centered at \SI{3217}{\keV} is consistent with the Hoyle
    state decaying via the first-excited state.}
  \label{fig:coin_spec}
\end{figure}

To obtain a coincidence spectrum, a gate was placed on the relative time between the two
$\gamma$ rays
and on the energy of the \mfex transition.
The widths of these gates were chosen to minimize any systematic effects.%
.

The coincidence spectrum is given in Fig.~\ref{fig:coin_spec}, where a clear peak centered at
\SI[parse-numbers=false]{3216.9_{\pm0.4 (stat)}^{\pm 0.7 (sys)} }{\keV} is visible. This is
consistent with a cascade decay of the Hoyle state via the first-excited level. The peak was
fitted with the same functional form as in the previous section, but the parameters for the
skewed Gaussian are determined from 
peaks I-III in Fig.~\ref{fig:full_spec}. Peak I-III originate from \ce{^{56}Mn} and
\ce{^{56}Co} produced in beam by reactions with \ce{Ti}. The area of the peak, determined from
the fit, is $N_{\gamma \gamma} = \num{58(9)}$.

\subsection{Efficiency}

The relative efficiency was determined using the standard calibration sources \ce{^{152}Eu} and
\ce{^{56}Co} mounted at the target position. This provides calibration points, both at low
energy and in the important \mSI{3}{\MeV} region. The absolute efficiency was calculated using
the coincidence method, including a correction for random coincidence events, for both a
\ce{^{60}Co} source and \ce{^{24}Mg}, which was produced by in-beam reactions
\cite{Siegbahn}. From this procedure, the absolute efficiency at \SI{3217}{\keV} was determined
to be $\epsilon_{3.21} = \num{2.94(2)} \%$.


\subsection{Angular correlation}
\begin{figure}[t]
  \centering
  \includegraphics[width=\mwidth, clip=true, trim=0 0 0 5mm]{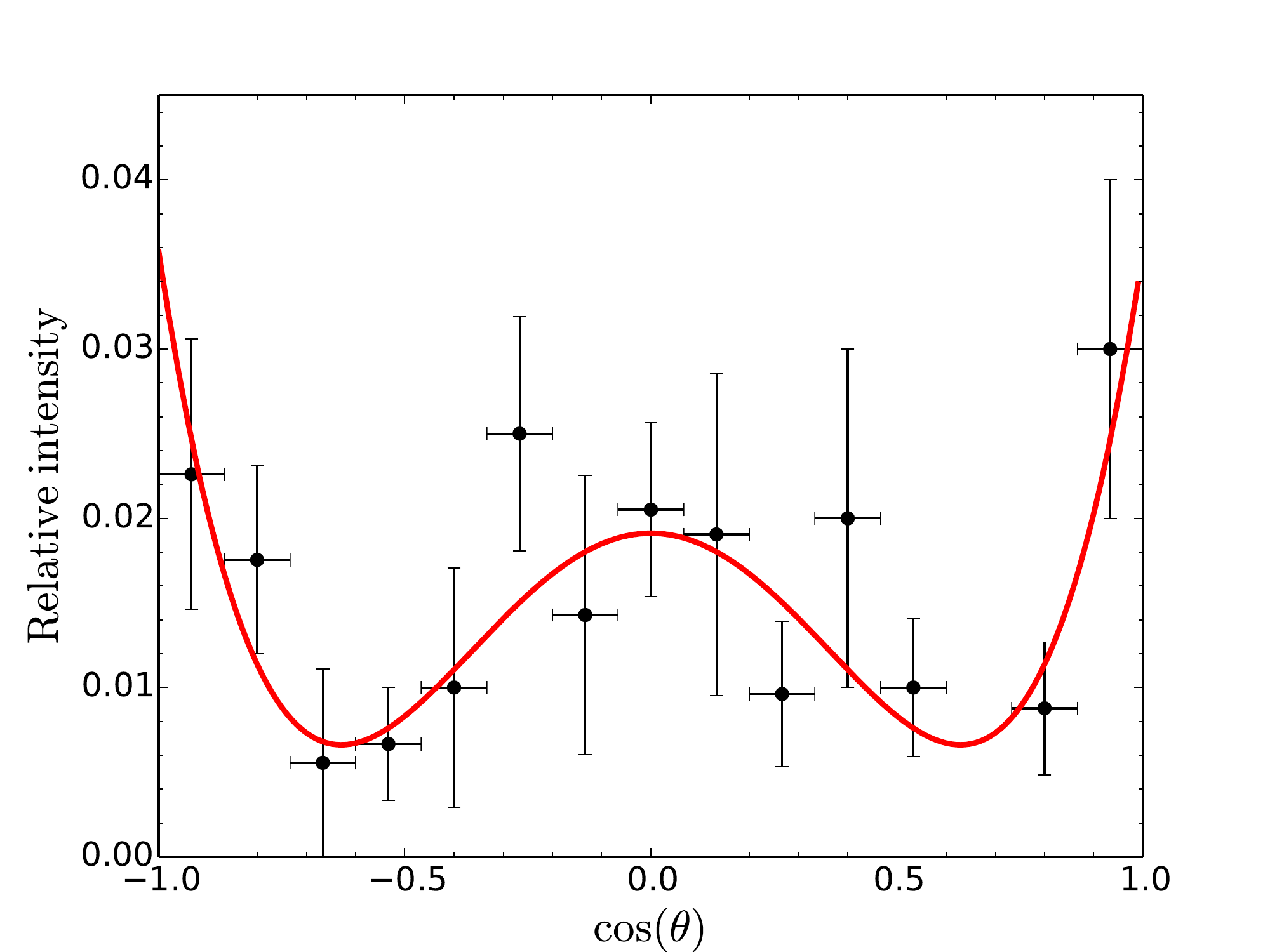}
  \caption{(Color online) Angular correlation of the Hoyle state $\gamma$
    cascade corrected for the geometric efficiency (number of detector pairs with
    a given angle between them). The solid line shown is the best fit to
    equation~\eqref{eq:angular}.}
  \label{fig:angular}
\end{figure}

Due to the excellent angular coverage of Gammasphere, it is possible to measure the angular
correlation of the two $\gamma$ rays,
which had not been measured previously.  Using the gates described above and in
addition requiring the energy of the second $\gamma$ ray
to be within \SI{10}{\keV} of \SI{3217}{\keV}, it is possible to extract the true coincidence events
plus some background. The shape of the background was determined by gating outside the peak,
and was found to be flat.

The angular correlation, corrected for the geometric efficiency (number of detector pairs with
a given angle between them), is shown in
Fig.~\ref{fig:angular}, together with the best fit to the equation
\begin{equation}
  \label{eq:angular}
  W(\theta) = k\left[1+a_{2}\cos^2(\theta)+a_{4}\cos^4(\theta)\right],
\end{equation}
where $\theta$ is the angle between the two $\gamma$ rays. 
The result of the fit is $a_{2} = \num{-3.3(7)}$
and $a_{4} = \num{4.2(9)}$,
which is consistent with the theoretical expectations $a_{2} = -3$
and $a_{4} = 4$ for a $0\rightarrow2\rightarrow0$ cascade \cite{Brady1950}.

With the theoretical angular correlation confirmed, it can be used to estimate the correction
factor $C_{\theta}$
from eq.~\eqref{eq:br}. This is done with a simple Monte Carlo simulation of the detector
setup, which gives $C_{\theta} = \num{1.00(1)}$,
as was expected from the large angular coverage by Gammasphere.

\subsection{Extraction of branching ratio}

The property directly measured in this experiment is the product of the relative $\gamma$ width
and the $\beta$ feeding of the Hoyle state
\begin{equation}
  \label{eq:product}
  \BR(7.65) \frac{\Gamma_{\gamma}}{\Gamma} = 
  \num{2.6(4)E-4}.
\end{equation}
Inserting the calculated value for the relative $\gamma$ width into equation~\eqref{eq:product} gives
\begin{equation}
  \label{eq:result}
  \BR(7.65) = \SI{0.64(11)}{\%},
\end{equation}
which is clearly inconsistent with the previous literature value of \litvalue
\cite{Selove1990,Hyldegaard2009b}, but agrees with that of \hylvalue found in Ref.
\cite{Hyldegaard2009}. Therefore, the feeding of the Hoyle state from \Boron is roughly a factor of $2$
smaller than indicated by Refs. \cite{Selove1990,Hyldegaard2009b}.

\section{Discussion}

The branching ratio from \Boron and \Nitro to the Hoyle state is a sensitive way to probe the
clustering of this state, as the $\beta$-decay
matrix element to the pure $3\alpha$
system is exactly zero due to the Pauli principle \cite{Kanada-Enyo2007}. The fact that $\beta$
decay is possible means that the Hoyle state must contain some $\alpha$-cluster
breaking component. Theoretically, this is obtained by mixing shell-model-like states with
cluster states as it is done; e.g; in Fermionic Molecular Dynamics (FMD)
\cite{Roth2004,Chernykh2007} and Antisymmetrized Molecular Dynamics (AMD) approach
\cite{Kanada-Enyo2007}. Alpha-cluster breaking was explicitly investigated in
Ref. \cite{Suhara2015} using a hybrid shell/cluster model, where it was found that the
spin-orbit force significantly changes the excited $0^{+}$ states.

Here, we compute the $\log ft$
value, which can be directly compared with these models. The available phase space ($f$ factor)
for $\beta$ decay from the ground state of \Boron to the Hoyle state was computed using the method in
\cite{Wilkinson1974}, with the excitation energy and half life from \cite{Selove1990}. With
this input our result is
\begin{equation}
  \label{eq:logft}
  \log ft = \num{4.50(7)}. 
\end{equation}
Due to the large change of the measured branching ratio compared to previous results
\cite{Selove1990}, the theoretical prediction of the AMD model, $\log ft = \num{4.3}$
\cite{Kanada-Enyo2007}, is no longer compatible with the experiment.

Hence, our branching ratio, together with the branching ratio for both \Boron and \Nitro from
Hyldegaard \etal \cite{Hyldegaard2009}, indicate that the $\alpha$
clustering of the Hoyle state is more pronounced than previously believed.

\section{Conclusion and outlook}

The $\beta$-decay branching ratio from \Boron to the second-excited state of \Carbon has been
measured using an array of high-purity Compton-suppressed germanium detectors. The branching
ratio was determined by counting the Hoyle state $\gamma$ decay,
and normalizing to the decay of the first-excited state. The result is \SI{0.64(11)}{\%},
consistent with the value found in \cite{Hyldegaard2009}, but is a factor $\unsim 2$
smaller than the previously-established value from \cite{Selove1990}. The updated branching was
used to compute $\log ft = \num{4.50(7)}$,
which is not consistent with latest results from AMD calculations \cite{Kanada-Enyo2007}. Our
results indicate that the clustering of the Hoyle state is more pronounced than previously
thought.

The angular correlation between the two photons emitted in the decay of the Hoyle state has also
been measured. The distribution was consistent with theoretical  expectations \cite{Brady1950}.

The errors on the present measurement are dominated by the uncertainty on the number of
coincidence events, which contributes \SI{91}{\%} of the total error, while \SI{6}{\%} and
\SI{2}{\%} come from the branching ratio to the first-excited state and the relative $\gamma$
width of the Hoyle state, respectively. Therefore, it is possible to make a $\unsim \SI{6}{\%}$
measurement of either the $\gamma$ width or the $\beta$-branching ratio by increasing statistics.

During the experiment, the beam current was limited to \SI{2}{p\nA} in order to minimize
neutron damage to Gammasphere. The main source for these neutrons was reactions with
titanium since the beam energy is above the Coulomb barrier. Exchanging titanium with hafnium
permits running with higher beam currents which, when combined with digital Gammasphere
\cite{Anderson2012}, would make it possible to accumulate sufficient statistics. Research into
production of such a target is ongoing.

\section{Acknowledgments}

\begin{acknowledgments}
  This work is supported by the U.S. Department of Energy, Office of Science, Office of Nuclear
  Physics, under Contracts No. DE-AC02-06CH11357 and No. DE-FG02-04ER41320 and Grant
  No. DE-FG02-94ER40834. This research used resources of ANL's ATLAS facility, which is a DOE
  Office of Science User facility.  OSK acknowledges support from the Villum Foundation. We
  also acknowledge financial support from the European Research Council under ERC starting
  grant LOBENA, No. 307447
\end{acknowledgments}

\bibliography{ANL1373-2}

\end{document}